\documentstyle[12pt]{article}
\input epsf.tex

\begin{document}

\baselineskip=15.5pt
\pagestyle{plain}
\setcounter{page}{1}

\renewcommand{\thefootnote}{\fnsymbol{footnote}}

\begin{titlepage}

\begin{flushright}
IASSNS-HEP-98/64\\
PUPT-1804\\
hep-th/9807080
\end{flushright}
\vfil

\begin{center}
{\huge Superconformal
Field Theory on Threebranes at a Calabi-Yau Singularity
}
\end{center}

\vfil
\def\Z{{\bf Z}}
\begin{center}
{\large Igor R.\ Klebanov\footnote{On leave from 
Joseph Henry Laboratories, Princeton University.}
and Edward Witten}\\
\vspace{1mm}
Institute for Advanced Study,
Olden Lane,
Princeton, New Jersey 08540, USA\\
\vspace{3mm}
\end{center}

\vfil

\begin{center}
{\large Abstract}
\end{center}

\noindent
Just as parallel threebranes on a smooth manifold are related to
string theory on $AdS_5\times {\bf S}^5$, parallel threebranes
near a conical singularity are related to string theory on $AdS_5\times X_5$,
for a suitable $X_5$.  For the example of the conifold singularity,
for which $X_5=(SU(2)\times SU(2))/U(1)$, we argue that string theory
on $AdS_5\times X_5$ can be described by a certain ${\cal N}=1$ supersymmetric
gauge theory which we describe in detail.

\vfil
\begin{flushleft}
July 1998
\end{flushleft}
\end{titlepage}
\newpage

\renewcommand{\thefootnote}{\arabic{footnote}}
\setcounter{footnote}{0}

\def\R{{\bf R}}
\def\Tr{{\rm Tr}}

\newcommand{\grad}{\nabla}
\newcommand{\tr}{\mathop{\rm tr}}
\newcommand{\half}{{1\over 2}}
\newcommand{\third}{{1\over 3}}
\newcommand{\be}{\begin{equation}}
\newcommand{\ee}{\end{equation}}
\newcommand{\bea}{\begin{eqnarray}}
\newcommand{\eea}{\end{eqnarray}}

\newcommand{\dint}[2]{\int\limits_{#1}^{#2}}
\newcommand{\D}{\displaystyle}
\newcommand{\PDT}[1]{\frac{\partial #1}{\partial t}}
\newcommand{\PD}{\partial}
\newcommand{\tw}{\tilde{w}}
\newcommand{\tg}{\tilde{g}}
\newcommand{\newcaption}[1]{\centerline{\parbox{6in}{\caption{#1}}}}
\def\href#1#2{#2}  

\def \ci {\cite}
\def \foot {\footnote}
\def \bi{\bibitem}
\newcommand{\rf}[1]{(\ref{#1})}
\def \del{\partial}
\def \m {\mu}
\def \n {\nu} 
\def \g {\gamma}
\def \G {\Gamma}
\def \a {\alpha}
\def \ov {\over}
\def \la {\label}
\def \ep {\epsilon}
\def \d {\delta}
\def \k {\kappa}
\def \p {\phi}
\def \ha {\textstyle{1\ov 2}}

\def\np {  {\em Nucl. Phys.} }
\def \pl { {\em Phys. Lett.} }
\def \mpl { Mod. Phys. Lett. }
\def \prl { Phys. Rev. Lett. }
\def \pr  { {\em Phys. Rev.} }
\def \cqg { Class. Quantum Grav.}
\def \jmp { Journ. Math. Phys. }
\def\ap { Ann. Phys. }
\def \ijmp { Int. J. Mod. Phys. }

\section{Introduction}

Recently, Maldacena argued that the `t Hooft
large $N$ limit \cite{GT} of
${\cal N}=4$ $SU(N)$ gauge theory is related to Type IIB
strings on $AdS_5\times {\bf S}^5$ \cite{jthroat}.
(For earlier work
on relations between large $N$ gauge theories and strings or supergravity,
see \cite{kleb,gukt,gkThree,Sasha}.)
This correspondence was sharpened in 
\cite{US,EW},
where it was shown how to calculate the correlation functions of  
gauge theory operators from the response of the Type IIB theory
on $AdS_5\times {\bf S}^5$ to boundary conditions. 

An interesting generalization of this duality between
gauge theory and strings is to consider other backgrounds of
Type IIB theory of the form $AdS_5\times X_5$ where $X_5$ is a
five-dimensional Einstein manifold bearing five-form flux. 
The arguments given in \cite{jthroat,EW}
indicate that these backgrounds are related to four-dimensional
conformal field theories. In general, these theories are
different from the ${\cal N}=4$ $SU(N)$ gauge theory.
This is obvious from the fact that only ${\bf S}^5$ preserves the maximal
number of supersymmetries (namely 32) while other Einstein manifolds
lead to reduced supersymmetry. 
In the early days of Kaluza-Klein supergravity, Romans gave a
partial list of five-dimensional Einstein manifolds together with their
isometries and the degree of supersymmetry \cite{Romans}.
It is of obvious interest to find a
field theoretic interpretation of the Romans compactifications,
and in this paper we report on some progress in this direction.

Most of the Einstein manifolds considered by Romans preserve no
supersymmetry, which makes construction of the field theory
difficult. Instead, we will focus on the cases with 
some unbroken supersymmetry.  In one particular case, we will identify
string theory on $AdS_5\times X_5$ with a field theory.
This is the case that  $X_5$ is
a homogeneous space $T^{1,1}=(SU(2)\times SU(2))/U(1)$, with the $U(1)$
being a diagonal subgroup of the maximal torus of $SU(2)\times SU(2)$.
(If 
$\sigma^{L,R}_i$ are the generators of the left and right
$SU(2)$'s, then the $U(1)$  is generated by
$\sigma^L_3 + \sigma^R_3$.) According to the counting of supersymmetries
in \cite{Romans},  
this compactification should be dual to an
${\cal N}=1$ superconformal field theory in four dimensions.

We will find that the construction of the field theory is
greatly facilitated by the observation that it is the infrared limit
of the world volume theory on coincident 
Dirichlet threebranes \cite{brane,Witten}
placed at a conical singularity of
a non-compact Calabi-Yau threefold. Thus, we are dealing with
a special case
of a connection between compactification on Einstein manifolds
and the metric of threebranes placed at
Calabi-Yau singularities, which was recently
pointed out in \cite{Kehagias}. The subject has been investigated
independently of the present work in \cite{Morr}.

We will also find some interesting relations with
theories on D3-branes placed at orbifold singularities, which have been
extensively studied  \cite{dm,lnv,bkv} and in particular
related to $AdS$ 
compactifications  \cite{ks}.
In general, these theories are related to Type IIB on
$AdS_5\times {\bf S}^5/\Gamma$ where $\Gamma$ is a discrete subgroup of 
$SU(4)$. It turns out that the coset space 
$T^{1,1}=(SU(2)\times SU(2))/U(1)$
may be obtained by taking $\Gamma={\bf Z}_2$, embedded in $SU(4)$ so as to
preserve ${\cal N}=2$ supersymmetry, and
blowing up the fixed circle of ${\bf S}^5/{\bf Z}_2$ (an operation that
breaks ${\cal N}=2$ to ${\cal N}=1$).
On the field theory side, the ${\cal N}=2$ superconformal theory
corresponding to ${\bf S}^5/{\bf Z}_2$ flows to the ${\cal N}=1$
IR fixed point corresponding to $T^{1,1}$. The necessary relevant
perturbation of the superpotential is odd under the
${\bf Z}_2$ and, therefore,
corresponds to a blow-up mode of the orbifold \cite{Han,Guk}.

\section{Branes at Conical Singularities and Einstein Spaces}

We will consider parallel branes near a conical singularity.
By a conical singularity on an  $n$-dimensional manifold $Y_n$, 
we mean a point (which we will label as $r=0$) near which
the metric can locally be put in the form
\be
\label{metric}
h_{mn} dx^m dx^n = dr^2 + r^2 g_{ij} dx^i dx^j\ , \qquad (i,j= 1, \ldots, n-1)
\ .
\ee
Here $g_{ij}$ is a metric on an $n-1$-dimensional manifold $X_{n-1}$;
the point $r=0$ is a singularity unless $X_{n-1}$ is a round sphere.
The basic property of this metric, which makes it ``conelike,'' is that
there is a group of diffeomorphisms of $Y_n$ that rescale the metric.
This group is $r\to tr$ with $t>0$; the group is thus isomorphic
to ${\bf R}^*_+$ (the multiplicative group of positive real numbers).
We call $Y_n$  a cone over $X_{n-1}$; $X_{n-1}$ is obtained if one deletes
the singularity at $r=0$ and divides by ${\bf R}^*_+$.

For $n>2$, the condition that $Y_n$ is
Ricci-flat is that that $X_{n-1}$ is an Einstein manifold of positive
curvature.  (For example, if $X_{n-1}$ is a round sphere, then $Y_n$
 is flat, not just Ricci-flat.)
In fact, 
by a conformal transformation the metric on $Y_n$ can be brought to the form
\be
\label{confmetric}
\hat h_{mn} dx^m dx^n = d\phi^2 + g_{ij} dx^i dx^j\ , \qquad 
\phi= \ln r
\ .
\ee
If $h_{mn}$ is a Ricci flat metric, then by applying the conformal
transformation to the Ricci tensor we find that $g_{ij}$ is an
Einstein metric,
\be
\label{curvnorm}
R_{ij} = (n-2) g_{ij}
\ .
\ee

\bigskip
\noindent
{\it Threebranes}

We will mainly consider the case that $n=6$.  
We take space-time to be
 $M_4\times Y_6$, with $M_4$ being four-dimensional Minkowski
space and $Y_6$ as above.  We consider $N$ parallel D3-branes
on $M_4\times P$, with $P$ the singularity of $Y_6$.
The
resulting ten-dimensional space-time has the metric \cite{Kehagias,afm}
\be
\label{threemetric}
   ds^2 = 
H^{-1/2}(r)
    \left[ - dt^2 + d\vec{x}^2 \right] +
H^{1/2}(r)
    \left[ dr^2 + r^2 g_{ij} dx^i dx^j \right] \ ,
\end{equation}
where 
$$   H(r)  = 1 + {L^4 \over r^4} \ , \qquad \  \ 
L^4= 4\pi g_s N (\alpha')^2 
\ .
$$
The near-horizon ($r\rightarrow 0$) limit of the geometry 
is $AdS_5\times X_5$ and, as we have shown,
$X_5$ is an Einstein manifold.
Thus, in the spirit of \cite{jthroat}, we may identify
the field theory on the D3-branes at a conical singularity
as the dual of Type IIB string theory on $AdS_5\times X_5$.

The above considerations facilitate the counting of
unbroken supersymmetries.
The Killing spinor equation in the metric $h_{mn}$ is
\be
(\partial_m+ {1\over 4} \omega_{mab} \Gamma^{ab} )\eta =0
\ .
\ee
Evaluating this in the metric (\ref{metric}), we find
\cite{Kehagias}
\be
(\partial_i+ {1\over 4} \omega_{ijk} \Gamma^{jk}+
{1\over 2} \Gamma_i^r)\eta =0
\ ,
\ee
which is equivalent to the $X_5$ part of
the  Killing spinor equation in Type IIB compactification on
$AdS_5\times X_5$, including the effects of the five-form field
 strength (which contributes the $\Gamma_i^r$ term).
Here $ \Gamma_i^r = \Gamma_{is} n^s$ where $n^s$ is the unit vector
in the radial direction. 
Thus, the number of unbroken supersymmetries on $X_5$ is the same as the
number of unbroken supersymmetries on the six-dimensional cone.
If the cone is a manifold of $SU(3)$ holonomy
(a Calabi-Yau threefold), then there are eight unbroken
supersymmetries -- four for left-movers and four for right-movers. 
These cases correspond to ${\cal N}=1$
superconformal field theories in 4 dimensions, i.e.
we may construct such field theories 
as the infrared limits of the theories on D3-branes placed at 
Calabi-Yau singularities.\foot{Recall that an ${\cal N}=1$ superconformal
field theory has eight fermionic symmetries -- four ordinary
supersymmetries and four superconformal symmetries.}
If, however, the cone has
$SU(2)$ holonomy, then by placing D3-branes at the singularity
we obtain an ${\cal N}=2$ superconformal field theory.  

This point
of view also helps in identifying which symmetries will be $R$-symmetries
of the superconformal field theory.  For example, 
to identify the R symmetries of string theory on $AdS_5\times X_5$
we work on a Calabi-Yau threefold $Y_6$ which is a cone over $X_5$.  
Let $\Omega$ be a holomorphic three-form
such that $-i\Omega\wedge \overline\Omega$ is the volume form of
$Y_6$.  Then the chiral superspace volume form $d^2\theta$ (the $\theta$'s 
being
fermionic coordinates of positive chirality) transforms like $\Omega$,
so symmetries are $R$-symmetries precisely if they act non-trivially
on $\Omega$.

\bigskip
\noindent{\it An Example}

For the special case where the five-dimensional Einstein manifold $X_5$
is $T^{1,1}=(SU(2)\times SU(2))/U(1)$, 
the Calabi-Yau threefold turns out to be particularly
simple. 
Since $T^{1,1}$ has $SU(2)\times SU(2)\times 
U(1)=SO(4)\times U(1)$ symmetry, we look for an isolated Calabi-Yau
singularity with that symmetry.  A fairly obvious candidate is
the ``conifold'' \cite{cgh,cd}
which for our purposes is the complex manifold $C$
\be\label{iconifold}
z_1^2+z_2^2+z_3^2+z_4^2=0
\ee
with an ``ordinary double point'' singularity at $z_i=0$.
The $z_i$ transform in the four-dimensional representation of 
$SO(4)$, and have ``charge one'' relative to the $U(1)$.
The holomorphic three-form is
\be \label{jconifold}
\Omega={dz_2\wedge dz_3\wedge dz_4\over z_1},
\ee
and has ``charge two'' under the $U(1)$, which will therefore be an 
$R$-symmetry group.

This manifold $C$ is a cone because the equation defining it
transforms with
definite weight under $z_i\to tz_i$, with $t\in {\bf C}^*$.  This
${\bf C}^*$ action is in fact the complexification of the $U(1)$ $R$-symmetry
group noted above.  If we restrict $t$ to be real and positive, we get
a group ${\bf R}^*_+$ of scalings under which the Calabi-Yau
metric that we will find momentarily transforms homogeneously, just like
the holomorphic three-form.

To identify $X_5$ topologically, we note that after omitting the
singularity at the origin, dividing $C$ 
by $\R^*_+$ is equivalent to intersecting
it with the unit sphere
\be\label{unitsphere}
|z_1|^2+|z_2|^2+|z_3|^2+|z_4|^2=1.
\ee
The group $SO(4)$ acts transitively on this intersection.  Any
given point on the intersection, such as $Q:(z_1,z_2,z_3,z_4)=(1/\sqrt 2,
i/\sqrt 2,0,0)$, is invariant under only a single $U(1)\subset SO(4)$
(for instance, $Q$ is invariant under the subgroup $U(1)=SO(2)$ that rotates
$z_3$ and $z_4$), so $X_5=SO(4)/U(1)=(SU(2)\times SU(2))/U(1)$
\cite{cd}.

A Calabi-Yau metric on $C$ can be written quite explicitly \cite{cd}.
One can describe any $SO(4)$-invariant
Kahler metric on $C$ by an $SO(4)$-invariant Kahler potential
$K$.  The most general possibility is that $K(z_1,z_2,z_3,z_4)$ must be a 
function only of $\sum_i\bar z_iz_i$.
To get a conical metric -- which scales homogeneously
under $z_i\to tz_i$ for $t\in {\bf R}^*_+$ -- we must take
$K=(\sum_i\bar z_iz_i)^\gamma$ for some exponent $\gamma$.
The Kahler form $\omega=-idz_id\overline z_j(\partial^2K/\partial z_i\partial
\overline z_j)$ scales as $\omega\to t^{2\gamma}\omega$.  We determine
$\gamma$  by asking that the metric should be a Calabi-Yau metric;
this is equivalent to the requirement that $\omega\wedge\omega\wedge \omega
=-i\Omega\wedge \overline \Omega$.  Since $\Omega$ transforms
as $\Omega\to t^2\Omega$, we get $\gamma=2/3$.

Starting from this description, to exhibit
the metric in the standard conical form of (\ref{metric}), one reparametrizes
${\bf R}^*_+$ by a new scaling variable $\tilde t=t^{2/3}$, so that
the Kahler form $\omega$ (or equivalently the Kahler metric) scales
as $\omega \to \tilde t^2 \omega$.  Then we introduce a radial variable
$r=(\sum_i\bar z_iz_i)^{1/3}$, which transforms as $r\to \tilde t r$.
In terms of $r$ plus a set of angular variables (invariant under scaling)
the metric takes the form (\ref{metric}).

The angular part of the
metric  has been described as follows \cite{Page,cd}.
$X_5=T^{1,1}$ is a $U(1)$ bundle over ${\bf S}^2\times {\bf S}^2$.
We choose the coordinates $(\theta_1,\phi_1)$ and $(\theta_2,\phi_2)$
to parametrize the two spheres in a conventional way, while
the angle $\psi \in [0, 4\pi)$ parametrizes the $U(1)$  fiber.
Then the metric may be written as
\be
\label{tmetric}
ds^2 = a (d\psi + \cos \theta_1 d\phi_1+ \cos \theta_2 d\phi_2)^2+
b \sum_{i=1}^2 \left [
d\theta_i^2 + \sin^2\theta_i d\phi_i^2 \right ]
\ .
\ee
If we choose $a=1/9$ and $b=1/6$,
we obtain an Einstein metric with $R_{ij} = 4 g_{ij}$.  
It can be obtained as the angular part of the conical
Calabi-Yau metric described above \cite{cd}.

\def\RP{{\bf RP}}

\bigskip
\noindent
{\it Description As A Quotient}

To construct a field theory that describes
threebranes at a conifold singularity,
it is very helpful to have a description of the conifold
as a quotient.  (This description has been used \cite{Wit} in analyzing
the structure of Kahler moduli space in Calabi-Yau compactification.)
After an obvious linear change of variables, we can describe
$C$ by the equation
\be
\label{conifold}
z_1 z_2 - z_3 z_4 =0
\ .
\ee
This equation can be ``solved'' by writing
\be
\label{param}
z_1 = A_1 B_1\ ,\qquad
z_2 = A_2 B_2\ ,\qquad
z_3 = A_1 B_2\ ,\qquad
z_4 = A_2 B_1\ .\qquad
\ee
Note that we obtain the  same $z_i$ if we transform the $A$'s and $B$'s by
\be
A_k \rightarrow \lambda A_k\ ,\qquad
B_l \rightarrow \lambda^{-1}B_l \ ,\qquad k,l=1,2
\ee
with $\lambda\in {\bf C}^*$.
The $SO(4)=SU(2)\times SU(2)$ symmetry of the conifold is easy to describe
in this formulation: one $SU(2)$ acts on the $A_i$, and one on the $B_j$.
If we write $\lambda=se^{i\alpha}$, with $s$ real and positive
and $\alpha$ real, then, away from the singular point $z_i=0$, $s$
can be selected to set
\be\label{selset}
|A_1|^2+|A_2|^2=|B_2|^2+|B_2|^2.
\ee
The conifold is obtained by further dividing by $U(1)$:
\be \label{condi}
A_k\rightarrow e^{i\alpha}A_k\ , \qquad
    B_l\rightarrow e^{-i\alpha}B_l\ . 
    \ee
To identify the angular manifold $X_5$ from this point of view,
note that we can divide by the scaling $z_i\to sz_i$ (with real positive
$s$) by setting $|A_1|^2+|A_2|^2=B_1|^2+|B_2|^2=1$.  At this
point we are on ${\bf S}^3\times {\bf S}^3=SU(2)\times SU(2)$.
Then dividing by (\ref{condi}) gives us back $X_5=(SU(2)\times SU(2))/U(1)$.

Our goal in the next section
will be to find the  ${\cal N}=1$ superconformal field theory
which is dual to the Type IIB theory
compactified on $AdS_5\times T^{1,1}$.
We will think of this theory as the infrared limit of the theory on
$N$ coincident D3 branes placed at the conical singularity of
$M_4\times C$.

\section{Construction of the Field Theory}

Our construction of the ${\cal N}=1$
superconformal field theory on the D3-branes at
the conical singularity of $M_4\times C$ will be guided by
the parametrization of the conifold $C$ in terms of $A_1$,
$A_2$, $B_1$, $B_2$ given in (\ref{param}).
We consider a $U(1)$ gauge theory with ${\cal N}=1$ supersymmetry,
and  introduce $A_k$ and $B_l$, $k,l=1,2$ as  chiral superfields of charges
$1$ and $-1$ respectively.  The $D$ auxiliary field of the $U(1)$
vector multiplet is given by
\be \label{dfield}
D=|A_1|^2+ |A_2|^2 - |B_1|^2- |B_2|^2-\zeta,
\ee
with $\zeta $ the coefficient of the Fayet-Iliopoulos term in the Lagrangian.
The moduli space of vacua is found by setting $D$ to zero and dividing
by the gauge group $U(1)$.  If we set $\zeta=0$, then the condition
$D=0$ is the condition (\ref{selset}), and dividing by the gauge group
is the equivalence relation (\ref{condi}). 
So the moduli space of vacua
is the conifold $C$.
For $\zeta\not= 0$, one gets instead a resolution
of the conifold singularity, in fact two different ``small resolutions''
depending on the sign of $\zeta$.  (The ``flop'' between them at $\zeta=0$
is a prototype of topology change in Calabi-Yau sigma models.)
In the present paper, we wish to study threebranes on the conifold; so 
we set $\zeta=0$.

To describe parallel threebranes on $M_4\times C$, we really should
introduce a second $U(1)$, which will be the unbroken $U(1)$ on the threebrane
worldvolume.  
The gauge group is then $U(1)\times U(1)$,
and the chiral multiplets $A_i$ and $B_j$ have respective charges
$(1,-1)$ and $(-1,1)$. 
All chiral multiplets are neutral under the diagonal $U(1)$, which
thus decouples and is the expected free $U(1)$ gauge multiplet on
the threebrane worldvolume.  Modulo this free $U(1)$, the model is equivalent
to the one analyzed in the last paragraph.  The chiral multiplets
describe threebrane motion on $C$.  The model can thus be considered
to describe the low energy behavior of a threebrane on $M_4\times C$
whose worldvolume in a configuration of minimum energy
is of the form $M_4\times Q$, where $Q$ is a point in $C$ determined
by a choice of vacuum of the field theory.

\def\bar{\overline}
\def\N{{\rm N}}
Now we want to generalize to the case of $N$ parallel threebranes.
The natural guess is that the gauge group should be 
$U(N)\times U(N)$
gauge theory with  chiral  fields $A_k, \,k=1,2$ transforming now in the
$(\N,\bar \N)$ representation and $B_l,\,l=1,2$ 
transforming in the
$(\bar \N, \N)$ representation.  A renormalizable superpotential is not
possible, so as a first guess we suppose that the superpotential vanishes.
We will think of $A_k$ and $B_l$ as $N\times N$ matrices.
By assuming that the matrices $A_k,B_l$ are (in some basis)
diagonal, with distinct eigenvalues, one finds a family of vacua
parametrized by the positions of  $N$ threebranes at distinct
points on the conifold.   The gauge group is broken down to $U(1)^N$,
one factor of $U(1)$ for each threebrane.  This is an encouraging
sign, but it cannot be the whole story, since the vacua just described
have massless charged chiral multiplets that should not be present
in a theory describing $N$ threebranes at generic smooth points.

To proceed, we must introduce a superpotential that will give mass to the
unwanted massless multiplets.  The model without
the superpotential has an $SU(2)\times SU(2)$ symmetry,
 with one $SU(2)$
acting on the $A_k$ and one on the $B_l$.  There is also an anomaly-free
$U(1)$ R-symmetry, under which $A_k$ and $B_l$ both have charge $1/2$.
\foot{We define R-symmetries so that chiral superspace coordinates
have charge 1.  Gluino fields $\lambda$ hence have charge $1$,
so that the superspace field strength
${\cal W}=\lambda+\theta F+\dots$ transforms homogeneously.
For $A_k=a_k+\theta\psi_k+\dots$, $B_l=b_l+\theta\chi_l+\dots$
to have charge $1/2$, the fermion components $\psi$, $\chi$ have
charge $-1/2$.  The adjoint representation of $U(N)$ has $C_2=2N$,
while the fundamental has $C_2=1$.  With these values of the Casimirs
and R-charges, the anomalies in the R-symmetry cancel.}
This $SU(2)\times SU(2)\times U(1)_R$ symmetry is a symmetry of
the conifold $C$.  So the superpotential must preserve the 
symmetry.  The most general superpotential that does so is
\be\label{superpotential}
W={\lambda\over 2}
\epsilon^{ij}\epsilon^{kl}\Tr A_iB_kA_jB_l
\ee
for some constant $\lambda$.  Note that $W$ has $U(1)_R$ charge two,
the correct value for a superpotential.  

It is not hard to see that this superpotential does the right job
at the classical level. 
The model has a family of vacua in which the matrices $A_k, B_l$
are (in some basis) diagonal and otherwise generic.  The gauge group
is broken to $U(1)^N$.  The diagonal components of $A_k,B_l$ describe
the motion of $N$ threebranes on $C$.  The off-diagonal fields all
receive mass from the superpotential (plus Higgs mechanism).  So the low
energy theory is as desired.  

Another way to reach this conclusion is the following.  
The $U(N)\times U(N)$ gauge symmetry can be  broken down to a diagonal
$U(N)$ by letting $A_k$ and $B_l$ be multiples of the identity matrix.
In such a vacuum, one of the matrices $A_k, B_l$ can be Higgsed away.
The other three give three chiral superfields, say $X,Y,$ and $Z$, transforming
in the adjoint representation of the unbroken $U(N)$.  The superpotential
(\ref{superpotential}), 
after allowing for the expectation value of one
linear combination of the fields, reduces to a cubic expression $\Tr X[Y,Z]$.
The $U(N)$ gauge theory of three chiral superfields with that superpotential
 flows in the infrared to the ${\cal N}=4$ super Yang-Mills
theory that describes $N$ threebranes near a smooth point in $C$.  This
can be further Higgsed, if desired, to the $U(1)^N$ theory described
in the last paragraph.

\bigskip\noindent{\it Interpretation Of The Superpotential}

To formulate a precise conjecture, it remains to
 discuss the meaning of the superpotential
(\ref{superpotential}), which as a perturbation of free field theory
is unrenormalizable.  As in \cite{is},\cite{ls},\cite{kutasov}, 
the basic idea is to
first consider the theory with $\lambda=0$.  
We want to understand its infrared dynamics.  The $U(1)$ factors in the
gauge
group have positive beta function and will decouple in the infrared.
So in analyzing the infrared behavior, we replace the gauge group by
$SU(N)\times SU(N)$.

{}From the point of view
of either of the $SU(N)$ factors, this is an $SU(N)$
gauge theory with chiral multiplets transforming in $2N$ copies of
$\N\oplus \bar\N$.  That theory flows \cite{seiberg}
to a nontrivial infrared fixed point
at which the anomaly-free R-symmetry of the Lagrangian becomes the R-symmetry
in the superconformal algebra.  We assume that the same is true
for an $SU(N)\times SU(N)$ theory with two copies of $(\N,\bar\N)\oplus
(\bar \N,\N)$.
This theory has, microscopically, a unique anomaly-free
R-symmetry, analyzed above, under which the chiral superfields all
have charge $1/2$.  We expect this (as in the examples in \cite{seiberg})
to become the R-symmetry in the 
superconformal algebra of the infrared fixed point theory.  If so,
dimensions of chiral superfields at the nontrivial fixed point are determined
from their R-charges.  In particular, the superpotential $W$ of
(\ref{superpotential}) has R-charge 2, and hence is a marginal perturbation
of the infrared fixed point.  

One can be more precise here; using techniques in \cite{ls},
one can argue that the superpotential perturbation of this theory
is an exactly marginal operator and gives a line of 
fixed points.
As usual, this has to do with the high degree of symmetry
of the operator: it is invariant under $SU(2)\times SU(2)$.

Let us impose the conditions of conformal invariance on
the theory with the superpotential (\ref{superpotential}). From 
the vanishing of the exact beta function \cite{sv} 
we get the equation
\be
\label{exactbeta} 
3 C_2 (G) - \sum_{i=A_k, B_l} T(R_i) (1 - 2\gamma_i) = 0
\ .
\ee
Due to the $SU(2)\times SU(2)$ symmetry, the anomalous dimensions satisfy
\be
\gamma_{A_1}= \gamma_{A_2}\ ,\qquad 
\gamma_{B_1}= \gamma_{B_2}\ . 
\ee
The two anomalous dimensions, $\gamma_A$ and $\gamma_B$, 
are functions of the
gauge couplings $g_1$ and $g_2$, and the superpotential strength $\lambda$.
By applying
(\ref{exactbeta}) to either of the two $SU(N)$'s we find the condition 
\be \gamma_A (g_1, g_2,\lambda)+ \gamma_B (g_1, g_2,\lambda)+ 
{1\over 2}= 0\ .
\ee
Requiring the scale invariance of the superpotential leads to exactly
the same condition. So, we have one equation for three coupling constants,
and this gives a critical surface. The surface of
fixed points is generated by two exactly marginal operators,
the superpotential (\ref{superpotential}) and the difference
between the kinetic energies of the two $SU(N)$'s.
If we further impose a symmetry
under the interchange of the two gauge groups, then the fixed surface
degenerates into a fixed line.

The argument for exact marginality does not go through
if we consider a less symmetric superpotential.
For instance, we could consider a superpotential
\be h Tr (A_1 B_1)^2
\ .
\ee
This preserves a ${\bf Z}_2$ symmetry under the interchange of the two
gauge groups, and we will keep the gauge couplings equal,
$g_1=g_2=g$. Since the $SU(2)$'s are broken, 
there are two different anomalous dimensions: for $A_1 (B_1)$ and
for $A_2 (B_2)$. Now the vanishing of the beta functions gives
$$ \gamma_1 (g,h)+ \gamma_2 (g,h)+ {1\over 2}= 0\ ,
$$
while the scale invariance of the superpotential requires
$$ \gamma_1 (g,h) + {1\over 4}= 0\ .
$$
Now we find two equations for two different functions of $g$ and $h$,
and we do not expect a critical line to exist.

The conclusion that there are two and only two exactly marginal operators,
the superpotential (\ref{superpotential}) and the difference between
the two kinetic energies, is confirmed by considering
the special case of $N=2$, where our model is an 
$SU(2)\times SU(2)\sim SO(4)$ gauge theory  coupled to four vectors of
$SO(4)$. This theory has been studied by Intriligator and
Seiberg in \cite{is}. Here the superpotential (\ref{superpotential})
reduces to the
baryon operator introduced in \cite{is}, which was found to be
exactly marginal: under duality it goes into the difference between
the kinetic energies of the two $SU(2)$'s, which
is perturbatively marginal. This case was also emphasized by
Leigh and Strassler \cite{ls}
as an example of an operator that is 
irrelevant as a perturbation of the free fixed point but exactly
marginal as a perturbation of a nontrivial infrared fixed point.

We can now state a precise conjecture.  Type IIB string theory 
on $AdS_5\times T^{1,1}$, with $N$ units of Ramond-Ramond flux
on $T^{1,1}$, should be equivalent to the theory obtained
by starting with $SU(N)\times SU(N)$ gauge theory, with two copies
of $(\N,\bar\N)\oplus (\bar N,\N)$, flowing to an infrared fixed point,
and then perturbing by the superpotential (\ref{superpotential}).

\bigskip\noindent{\it Comparison Of R-Symmetries}

We will make several additional checks of this conjecture.

We begin with a more careful comparison of
the $U(1)$ R-symmetries 
of the conifold model and the field theory.  Consider the transformation
$z_i\to e^{i\phi}z_i$ of the conifold.  This transformation acts on
the holomorphic three-form  by $\Omega\to e^{2i\phi}\Omega$.  The
chiral superspace coordinates transform as $\Omega^{1/2} $ 
and thus as $e^{i\phi}$.  
\footnote{The reason for this
is that $\Omega$ can be written in terms of a covariantly constant spinor
$\eta$ of definite chirality
as $\Omega_{ijk}=\eta^T\Gamma_{ijk}\eta$, so $\Omega$ transforms
like $\eta^2$.  But supersymmetries are generated by covariantly constant
spinors, and so the supersymmetry generators transform like $\eta$.}

Now set $\phi=\pi$.  This gives an element of the R-symmetry group that
acts on the conifold as $z_i\to -z_i$ and on  the chiral superspace coordinates
$\theta$ by $\theta\to -\theta$.  Let us identify this transformation
in the gauge theory.  Since $A_k$ and $B_l$ have R-charge $1/2$, 
they transform under $\theta\to -\theta$ as $A_k\to iA_k$, 
$B_l\to iB_l$.
This agrees with expectations from the conifold since the conifold
coordinates $z_i$ are represented in the gauge theory as 
${\rm Tr} \,A_kB_l$
and thus indeed transform as $z_i\to -z_i$.  This is an interesting check
because it depends on the fact that $A_k$ and $B_l$ have 
R-charge $1/2$, a fact
that was deduced from considerations of anomaly cancellation that are
seemingly unrelated to the geometry of the conifold.

\bigskip\noindent{\it Global Structure Of The Symmetry Group}

We identified earlier  an $SU(2)\times SU(2)$ symmetry group of the gauge
theory, with one $SU(2)$ acting on $A_k$ and one on $B_l$.  
To be more precise,
the group that acts faithfully modulo gauge transformations
is $(SU(2)\times SU(2))/{\bf Z}_2$
where ${\bf Z}_2$ is the diagonal 
subgroup of the product of the centers of the
two $SU(2)$'s.  
In other words, the transformation $A_k\to -A_k,\,B_l\to B_l$
is equivalent to $A_k\to A_k, \,B_l\to -B_l$.  The reason for this
is that the gauge group $U(N)\times U(N)$ contains a $U(1)$ subgroup that
acts by $A_k\to e^{i\alpha}A_k$, 
$B_l\to e^{-i\alpha}B_l$.  Setting $\alpha=\pi$,
we get the gauge transformation
$A_k\to -A_k$, $B_l\to -B_l$, so that the transformations $A\to-A,B\to B$
and $A\to A, B\to -B$ are indeed gauge equivalent to each other.

The group acting faithfully is thus $(SU(2)\times SU(2))/{\bf Z}_2=SO(4)$.
But this is the global form of the corresponding symmetry group of the
conifold.  For indeed,
the conifold coordinates $z_i$, obeying $\sum_i z_i^2=0$, transform
in the vector representation of $SO(4)$.

\bigskip\noindent{\it Reflection}

Now we will analyze the discrete symmetries on the two sides.  First of all,
on the conifold side,  the $SO(4)$ symmetry group
extends to $O(4)$, as the conifold is invariant under $z_4\to -z_4$ with
other coordinates invariant.  
This transformation changes the sign of the holomorphic three-form $\Omega$,
so it is an R-symmetry, acting on the chiral 
superspace coordinates as $\theta\to
i\theta$.  

To compare to the gauge theory, we note that the reflection exchanges
the two factors in $SO(4)=(SU(2)\times SU(2))/{\bf Z}_2$.  So it must
exchange the $A$'s and $B$'s.  As $A$ transforms under $U(N)\times U(N)$
as $(\N,\bar\N)$ while $B$ transforms as $(\bar\N,\N)$,
 a transformation that exchanges $A$ and $B$ must
be accompanied by either (1) exchange of the two factors of the gauge group,
or (2) charge conjugation, that is an outer automorphism of each $U(N)$
that exchanges  the $\N$ and $\bar \N$ representations.

The reason that the two options exist is that, as we will see shortly,
the theory actually has global symmetries of each kind.  But for the moment,
we specify further that we want to examine a symmetry of the conifold that
acts by $z_4\to -z_4$ without reversal of the orientation of the string
worldsheet.  Charge conjugation is associated in $D$-brane theory with
such orientation reversal.  So the ``geometrical'' symmetry $z_4\to -z_4$
with no worldsheet orientation reversal corresponds to $A\leftrightarrow B$
together with exchange of the two factors in the gauge group.

It remains to understand, from the point of view of field theory, why
such a symmetry is an R-symmetry.   The point is that under 
$A_k\leftrightarrow
B_k$, the superpotential $W=\lambda\Tr (A_1B_1A_2B_2-A_1B_2A_2B_1)$ is odd.
So we must accompany the transformation described so far by an
additional transformation under which $W$ is odd.  Such a transformation,
if it exists, is not unique since one could always multiply by an ordinary
symmetry of the theory.  The missing transformation is uniquely determined
if one asks that it should leave fixed the lowest components of the superfields
$A_i$ and $B_j$.  The required 
transformation is the naive R-symmetry that
acts on chiral superspace coordinates by $\theta\to i\theta$, acts on gluinos
by $\lambda\to i\lambda$, leaves invariant the superfields $A$ and $B$,
and (therefore) acts on fermionic components $\psi$ of $A$ or $B$ by
$\psi\to -i\psi$.  We will call this transformation $\Upsilon$.
We note in particular that the fact that $\theta\to i\theta$ under $\Upsilon$
is in agreement with the $z_4\to -z_4$ symmetry of the conifold,
and that under such an R-transformation the superpotential is odd.

What remains is to show that the combined operation of exchanging $A$ and
$B$, exchanging the two factors in the gauge group, and acting with $\Upsilon$
has no anomaly.  There is a subtlety here that is relatively unfamiliar
since anomalies under global symmetries that act by outer automorphisms
of the gauge group are not often studied.  By itself (in addition to
being explicitly violated by the superpotential), $\Upsilon$ can sometimes
have an anomaly
in an instanton field.  In fact, in an instanton
field of the first $U(N)$ with instanton number $k$, 
the gluinos of the first $U(N)$ have $2Nk$ zero modes,
while $A$ and $B$ have $2Nk$ each, so the  path integral
measure transforms under $\Upsilon$ as $i^{2Nk}(-i)^{4Nk}=(-1)^{Nk}$,
and is invariant if and only if $Nk$ is even.  Since $\Upsilon$ is in any
case not a symmetry (being violated by the superpotential), whether 
$\Upsilon$ by itself has an anomaly is not of great physical relevance.
What we really want to know is whether there is an anomaly in $\Upsilon$
combined with the exchange of the two gauge group factors (and 
$A\leftrightarrow
B$).  For this, we look at a classical field configuration that is
invariant under exchange of the $U(N)$'s and the superfields, and study
how the fermion determinant in such a configuration transforms under 
$\Upsilon$.  For a classical configuration to be invariant under exchange
of the $U(N)$'s and the superfields, there must be equal instanton
numbers in the two $U(N)$'s.  The computation of the transformation
of the measure proceeds just as before, but now there is a factor of
$(-1)^{Nk}$ from each $U(N)$, so overall the measure is invariant.

\bigskip\noindent{\it Center Of $SL(2,{\bf Z})$}

One additional discrete symmetry of string theory on the conifold
should be compared to the gauge theory.

This is the element 
\be
\left(\matrix{-1 & 0 \cr 0 & -1 \cr}\right)
\ee
which generates the center of $SL(2,{\bf Z})$.  This symmetry, which we will
call $w$, acts trivially on the 
coupling (and theta angle) of Type IIB superstring theory, and is a symmetry
of Type IIB on the conifold as long as the $B$-fields vanish, as we have
so far assumed.  

The transformation $w$ is equivalent to $\Omega (-1)^{F_L}$, where $\Omega$
is the exchange of left and right movers on the string worldsheet, and
$(-1)^{F_L}$ multiplies left-moving worldsheet fermions by $-1$.  If
we write $Q_L$ and $Q_R$ for supersymmetries that come from left and right 
movers, then $\Omega$ acts by $Q_L\leftrightarrow Q_R$, and $(-1)^{F_L}$
by $Q_L\to -Q_L$, $Q_R\to Q_R$.  Hence $w$ acts by $Q_L\to Q_R$, $Q_R\to
-Q_L$.  Now let us consider what happens in the presence of parallel
threebranes whose world-volume is spanned by Minkowski coordinates
$x^0,\dots,x^3$.  
The unbroken supersymmetries are linear combinations of the form
$\epsilon_LQ_L+\epsilon_RQ_R$, where $\epsilon_R=\Gamma_{0123}\epsilon_L$.
Hence $w$, which acts by $\epsilon_L\to \epsilon_R$, $\epsilon_R\to -
\epsilon_L$, is equivalent to 
\be
w:\epsilon_L\to \Gamma_{0123}\epsilon_L.
\ee
$\Gamma_{0123}$ acts on spinors of positive chirality as $i$ and
on spinors of negative chirality as $-i$.  So $w$ (like the reflection
$z_4\to -z_4$ that we just analyzed) 
is an R-symmetry, acting on chiral superspace coordinates
by $\theta\to i\theta$.

Now let us look at what happens in field theory.  We expect 
$w=\Omega (-1)^{F_L}$ to act on the gauge group by charge conjugation.
The reason for this is that $\Omega$, the reversal of worldsheet orientation,
is the basic charge conjugation operation for open strings.  On the other
hand, $w$ commutes with $SU(2)\times SU(2)$ and so will {\it not}
exchange $A$ and $B$.  Since $A$ transforms as $(\N,\bar \N)$ and $B$
as $(\bar \N,\N)$, a charge conjugation operation that does not exchange
$A$ and $B$ must exchange the two factors in the gauge group.
So we identify $w$ with a transformation that exchanges the two factors
of the gauge group, in such a way that the $\N$ of the first
$U(N)$ is exchanged with the $\bar \N$ of the second (and $\bar \N$ of
the first is exchanged with $\N$ of the second) while mapping $A$ to $A$
and $B$ to $B$.

Let us suppress the $SU(2)$ internal symmetry index of $A$ and make
explicit the gauge indices.  Thus we write $A$ as a matrix $A^a{}_b$,
where $a$ labels the $\N$ of the first $U(N)$ and $b$ labels the
$\bar \N$ of the second $U(N)$.  The symmetry $w$ exchanges the two
types of indices, so we can regard it as an operation that maps $A$ to
its ``transpose'' $A^T$.  Under this operation of transposing all $A_k$
and $B_l$, the superpotential 
$W=\lambda {\rm Tr}\,(A_1B_1A_2B_2-A_1B_2A_2B_1)$
changes sign.  To compensate for this, we must include the action of
our friend $\Upsilon$, which we encountered in analyzing the reflection
$z_4 \to -z_4$.  Since $\Upsilon$ acts on chiral superspace coordinates
as $\theta\to i\theta$, $w$ acts in this way in the gauge theory,
just as we found in the underlying string theory.

\bigskip\noindent{\it Counting Of Moduli}

We will now make perhaps the most obvious comparison of all:
counting moduli on the two sides and discussing how they transform under
symmetries.

On the $AdS_5\times T^{1,1}$ side, one modulus is the $\tau$ parameter
of Type IIB.  An additional modulus arises
because topologically $T^{1,1}={\bf S}^2\times {\bf S}^3$.  (We give an 
explanation of this statement later.)  In particular, the second Betti
number of $T^{1,1}$ is 1. This means that, in Type IIB compactification
on $T^{1,1}$, the Ramond and NS $B$-fields each have a zero mode,
leading to two ``theta angles''  that appear in
labeling a choice of vacuum.   They combine under ${\cal N}=1$ supersymmetry
into a chiral superfield $\phi$, which parametrizes a complex torus.

On the gauge theory side, after decoupling the $U(1)$'s, one has the
renormalization scales $\Lambda_1$ and $\Lambda_2$ of the two $SU(N)$'s 
(as is standard in supersymmetric theories, we include the theta angles
in the definition of the $\Lambda_i$, so that they are complex-valued).
In addition, we have the coupling constant $\lambda$ that appears in the
superpotential (\ref{superpotential}). From
$\Lambda_1$, $\Lambda_2$ and $\lambda$, we can form two dimensionless
combinations $u_i=\lambda\Lambda_i$.  As we have explained earlier,
the gauge theory has
 a surface of fixed points parametrized
 by the $u_i$.

We propose that the two moduli $u_i$ of the gauge theory correspond
to the two moduli $\tau$ and $\phi$ in Type IIB on 
$AdS_5\times {\bf S}^5$.
As a check, let us look at the behavior under the symmetry $w$ that
generates the center of $SL(2,{\bf Z})$.  Under $w$, $\tau$ is even but
(as $w$ acts by $-1$ on both the Ramond and NS $B$-fields) $\phi$ is odd.
Now on the gauge theory side, we identified $w$ with an operation that
among other things exchanges the two $SU(N)$ factors of the gauge group. 
Under this transformation, clearly $u_1+u_2$ is even, but $u_1-u_2$ is odd.
So we propose to identify the even variable with the Type IIB coupling
$\tau$ and the odd variable with the theta angles $\phi$.

It is rather odd to associate a difference of gauge couplings $u_1-u_2$
with a torus-valued superfield $\phi$.  Such an identification has,
however \cite{lnv}, already been made in the case of a certain $AdS_5\times
{\bf S}^5/{\bf Z}_2$ orbifold.  As we will see next, this orbifold theory
can ``flow'' to the $AdS_5\times T^{1,1}$ vacuum considered in the present 
paper.
Thus the identification made in \cite{lnv} implies that the difference
of the  coupling parameters of the two $SU(N)$'s should indeed be identified
with the torus-valued parameter $\phi$.

\bigskip\noindent{\it Comparison To Orbifold Theory}

The duality that we have proposed can be checked in an interesting
way by comparing to a certain
$AdS_5\times {\bf S}^5/{\bf Z}_2$ background.  If ${\bf S}^5$ is described
by an equation
\be\label{beqnop}
\sum_{i=1}^6x_i^2=1,
\ee
with real variables $x_1,\dots, x_6$,
then the ${\bf Z}_2$ in question acts as $-1$ on four of the $x_i$ and
as $+1$ on the other two.  The importance of this choice is that this
particular
 ${\bf Z}_2$ orbifold of $AdS_5\times {\bf S}^5$
  has ${\cal N}=2$ superconformal symmetry.
Using orbifold results for branes \cite{dm}, this model has been 
identified \cite{ks} as an AdS dual of a $U(N)\times U(N)$ theory
with hypermultiplets transforming in
$(\N,\bar\N)\oplus (\bar \N,\N)$. From an ${\cal N}=1$ point of 
view, the hypermultiplets                
correspond to chiral multiplets $A_k,B_l$, $k,l=1,2$ in the 
$(\N,\bar \N)$ and $(\bar \N,\N)$  representations respectively.  
The model also contains, from an ${\cal N}=1$ point of view, chiral multiplets
$\Phi$ and $\tilde \Phi$ in the adjoint representations of the two
$U(N)$'s. 
The superpotential is
$$ g \Tr \Phi (A_1 B_1 + A_2 B_2) + g \Tr \tilde \Phi (B_1 A_1 + B_2 A_2)
\ .
$$
Now, let us add to the superpotential of this ${\bf Z}_2$ orbifold
a relevant term,
\be \label{relper}{m\over 2} (\Tr \Phi^2 - \Tr \tilde \Phi^2 )
\ .
\ee
It is straightforward to see what this does to the field theory.
We simply integrate out $\Phi$ and $\tilde \Phi$,
to find the superpotential
$$ {g^2\over 2m} \left [\Tr (A_1 B_1 A_2 B_2) - \Tr (B_1 A_1 B_2 A_2)
\right ]\ .
$$
This expression is familiar from (\ref{superpotential}), so the
${\bf Z}_2$ orbifold with relevant perturbation (\ref{relper}) 
apparently flows to the $T^{1,1}$ model associated with the conifold.
(However, this way of obtaining the $T^{1,1}$ model does not explain
all of its symmetries; the $T^{1,1}$ model has symmetries that arise
only at the endpoint of the renormalization group flow from 
${\bf S}^5/{\bf Z}_2$.)

\def\S{{\bf S}}
\def\Z{{\bf Z}}
Let us try to understand why this works from the point of view of the
geometry of ${\bf S}^5/{\bf Z}_2$.  The perturbation in (\ref{relper})
is odd under exchange of the two $U(N)$'s.  The exchange of the two $U(N)$'s
is the quantum symmetry of the $AdS_5\times {\bf S}^5/{\bf Z}_2$ orbifold
- -- the symmetry that acts as $-1$ on string states in the twisted sector
and $+1$ in the untwisted  sector.  So we must, as in \cite{Guk},
associate this perturbation with a twisted sector mode of string
theory on $AdS_5\times {\bf S}^5/\Z_2$.  The twisted sector mode
which is a relevant perturbation of the field theory is the blowup
of the orbifold singularity of $\S^5/\Z_2$.\foot{Note that in string
theory, blowup of a codimension four ${\bf R}^4/\Z_2$ singularity
is usually a marginal deformation, but in the present context, blowup
of the $\S^5/\Z_2$ singularity is a relevant deformation.  
This statement follows from a knowledge of the ${\cal N}=2$ super
Yang-Mills theory
dimensions but has not yet been explained from supergravity.}  

Let us consider the geometry produced by this blowup.  
First we recall the blowup (in the complex sense) of
a codimension four $\Z_2$ orbifold singularity ${\bf R}^4/\Z_2$.
Let ${\bf Z}_2$ act by sign change on four coordinates $x_1,\dots,x_4$ of
${\bf R}^4$.
Blowup of such a singularity replaces it by a copy of $\S^2$.
  One way to describe
the $\S^2$ is as follows.  The smooth part of ${\bf R}^4/\Z_2$, where the
$x_i$ are not all zero, maps
to ${\bf RP}^3$ by mapping $x_1,\dots,x_4$ to their image in real projective
space. 
One can identify ${\bf RP}^3$ with the $SO(3)$ group manifold.
$SO(3)$ can be projected to the two-sphere $\S^2=SO(3)/U(1)$, where
the $U(1)$ is a maximal torus of $SO(3)$.  
The resulting map of the smooth part of ${\bf R}^4/\Z_2$ to $\S^2$ can be made
completely explicit as follows.  Let $n_1=x_1+ix_2$, $n_2=x_3+ix_4$,
and map a point with coordinates $x_1,\dots,x_4$
to 
\be\label{bbb}
\vec b={(n,\vec\sigma n)\over (n,n)}
\ee
where $\vec\sigma$ are the Pauli matrices and the inner product
is such that $(n,n)=|n_1|^2+|n_2|^2$.
The map of ${\bf R}^4/\Z_2$ to $\S^2$ given by this formula
is ill-defined at the origin,
but the blowup precisely replaces the singular point at the origin
by a copy of $\S^2$ in such a way that the map is well-defined everywhere.

Now return to $\S^5/\Z_2$, which we describe as in (\ref{beqnop}).  
$\S^5/\Z_2$ is fibered over $\Z_2$, with a map that is given by precisely
the same formula as in (\ref{bbb}).  Just as in the ${\bf R}^4/\Z_2$ case,
the blowup renders this map well-defined even where $x_1,\dots,x_4$
all vanish.   The fiber of the map from $\S^5/\Z_2$ to $\S^2$ can
be determined by looking at the inverse image of any point, say the
point with $\vec b=(0,0,1)$.  The inverse image of this point is given
by $x_3=x_4=0$ and is a copy of $\S^3$.\footnote{There is a subtlety hidden
in this statement.   After setting $x_3=x_4=0$, we are left with 
$x_1,x_2,x_5,x_6$, with $x_1^2+x_2^2+x_5^2+x_6^2=1$; this looks like
$\S^3$, but we are to divide
by a sign reversal of $x_1,x_2$.  The quotient by that sign reversal
is a manifold, because the fixed points have codimension two, and is
a copy of ${\bf S}^3$.  One can prove this explicitly by writing $x_1=
\rho\cos \theta$, $x_2=\rho \sin\theta$ with $0\leq \theta\leq 2\pi$.
The sign change of $x_1,x_2$ amounts to $\theta\to \theta+\pi$; dividing
by this operation means that $\theta$ ranges from $0$ to $\pi$.
The effect of this can be undone topologically by just replacing
$\theta$ by $\theta'=2\theta$, which ranges from $0$ to $2\pi$.}  
  Hence the blowup of $\S^5/\Z_2$ is
an $\S^3$ bundle over $\S^2$.

What about $T^{1,1}=(SU(2)\times SU(2))/U(1)$?  By ``forgetting,'' say,
the second $SU(2)$, $T^{1,1}$ maps to $SU(2)/U(1)=\S^2$.  So $T^{1,1}$
is also a fiber bundle over $\S^2$.  The fiber is what was forgotten
in the map, namely the second $SU(2)$.  Since $SU(2)$ is isomorphic
topologically to $\S^3$, $T^{1,1}$ is also an $\S^3$ bundle over $\S^2$.

If we can show that these two $\S^3$ bundles are equivalent topologically,
we will get a new understanding of the relation between the $T^{1,1}$
and $\S^5/\Z_2$ models:  blowup of $\S^5/\Z_2$ has simply produced
$T^{1,1}$.  In fact, $\S^3$ bundles over $\S^2$ are classified
by $\pi_1(SO(4))=\Z_2$, so there are only two possibilities for
what our bundles might be.

Both $\S^3$ bundles we have met above
are special cases of the following more general construction.  Any circle
bundle $S$ over $\S^2$ can be built by starting with trivial bundles
over the upper and lower hemispheres of $\S^2$, and gluing them together
over the equator with the gluing function
\be\label{glufun}
\left(\matrix{\cos n\theta & \sin n\theta \cr -\sin n\theta & \cos n\theta\cr}
\right)
\ee
for some integer $n$.  Here $\theta$ is an angular variable on the equator
of $\S^2$, and we regard $S$ as an $SO(2)$ bundle.  It can be shown
that the integer $n$ equals the Euler class, or first Chern class, of
the circle bundle $S$.
Now let $S$ and $S'$ be circle bundles of first Chern class $n$ and
$n'$ respectively.  One can canonically make a three-sphere bundle
$W(S,S')$ whose fiber is the ``join'' of $S$ and $S'$; concretely,
if the fiber of $S$ is a circle $u^2+v^2=1$ and the fiber of $S'$ is
a circle $w^2+x^2=1$, then the fiber of $W(S,S')$ is the three-sphere
$u^2+v^2+w^2+x^2=1$.  $W(S,S')$  can be built starting with trivial
bundles over the upper and lower hemispheres of $\S^2$, and gluing
them on the equator with the gluing function
\be\label{moreglue}
\left(\matrix{\cos n\theta & \sin n\theta & 0 & 0 \cr
               -\sin n\theta & \cos n\theta & 0 & 0 \cr
                0 & 0 & \cos n'\theta & \sin n'\theta \cr
                 0 & 0 & -\sin n'\theta & \cos n'\theta \cr}\right).
\ee
This describes a trivial or non-trivial element of $\pi_1(SO(4))$ depending
on whether $n+n'$ is even or odd.  So the three-sphere bundle $W(S,S')$
is non-trivial if and only if $n+n'$ is odd.

As we will see, 
the blowup of $\S^5/\Z_2$ can be identified naturally as $W(S,S')$
with $n=2,$ $n'=0$, while $T^{1,1}$ can be identified naturally
as $W(S,S')$ with $n=n'=1$.  So these bundles are both topologically
trivial, isomorphic to the product $\S^2\times \S^3$.

To verify the claim about $\S^5/\Z_2$, we note that when $x_1,\dots ,x_4$
are not all zero, they determine (after dividing by $\Z_2$) an element
of $\RP^3$.  $\RP^3$, which can be regarded as the group manifold
of $SO(3)$, is a circle bundle over $\S^2=SO(3)/U(1)$ with Euler
class 2.  We let $S$ be this circle bundle over $\S^2$.  We let $S'$
be the trivial circle bundle, of $n'=0$, whose fiber is parametrized
by $x_5,x_6$ with $x_5^2+x_6^2=1$.  By reexamining the argument by which
we showed that $\S^5/\Z_2$ is a three-sphere bundle over $\S^2$, it can
be seen that the fiber is precisely $W(S,S')$ with the stated $S$ and $S'$.
The identification of $T^{1,1}$ with $W(S,S')$ where $n=n'=1$ is
more elementary.  It follows from noting that the second $SU(2)$,
which we ``forget'' to map $T^{1,1}=(SU(2)\times SU(2))/U(1)$ to
$\S^2$, can be written as a sphere $u^2+v^2+w^2+x^2=1$, where $U(1)$
acts by rotation of the $u-v$ plane together with rotation of the $w-x$ plane.

There is a somewhat 
shorter and, perhaps, more powerful argument for why
the blowup of the fixed circle of $\S^5/\Z_2$ gives $T^{1,1}$.
This argument relies on the fact that both sides have a $U(1)$
R-symmetry.  $\S^5/\Z_2$ and $T^{1,1}$ are $\S^1$
bundles over complex surfaces, say $B$ and $B'$.  $\S^5/\Z_2$ and
$T^{1,1}$ can be reconstructed from $B$ and $B'$ using the Calabi-Yau
condition, so it suffices to compare $B$ and $B'$.

Dividing $\S^5$ by $U(1)$ gives ${\bf CP}^2$; dividing
$\S^5/\Z_2$ by $U(1)$ gives ${\bf CP}^2/\Z_2$, where ${\bf CP}^2$ has
homogeneous coordinates $a_1,a_2,a_3$, and $\Z_2$ acts by sign reversal
on $a_1$, $a_2$.  ${\bf CP}^2/\Z_2$ has an orbifold singularity
at $a_1=a_2=0$; if this is deformed or blown up, we should get the quotient
of $T^{1,1}$ by $U(1)$, which is $\S^2\times \S^2$ or ${\bf CP}^1\times
{\bf CP}^1$.
We can describe ${\bf CP}^2/\Z_2$ in terms of the $\Z_2$-invariant
polynomials in $a_1,a_2,a_3$, which are generated by
\be
u_1= a_1^2\ , \quad u_2 = a_2^2\ , \quad u_3 = a_1 a_2\ , 
\ee
and $a_3$.
They are homogeneous coordinates for a weighted projective space
${\bf WCP}^{3}_{2,2,2,1}$, and obey
\be
\label{defineq}
u_1 u_2 - u_3^2 = 0
\ .
\ee
The weighted projective space is actually equivalent to an ordinary
projective space, for the following reason.  In the scaling
$(u_1,u_2,u_3,a_3)\to (\lambda^2u_1,\lambda^2u_2,\lambda^2u_3,\lambda a_3)$
by which  ${\bf WCP}^3_{2,2,2,1}$ is defined,
if we set $\lambda=-1$, we have simply $a_3\to -a_3$.  
So in dividing by this scaling to obtain the weighted projective
space ${\bf WCP}^3_{2,2,2,1}$, we among other things divide by $a_3\to -a_3$.
That step can be accomplished by restricting to the invariant functions,
which are generated by $u_1,u_2,u_3$, and $u_4=a_3^2$.
The $u_i$ all have the same weight (two), so are homogeneous coordinates
for an ordinary projective space ${\bf CP}^3$.  In this projective
space, ${\bf CP}^2/\Z_2$ is defined by the equation (\ref{defineq}).
To deform the singularity (which has the same topological effect
as blowing it up), we deform the equation to
\be
u_1 u_2 - u_3^2 + \epsilon u_4^2= 0
\ ,
\ee
which by an obvious linear change of variables can be brought to the
form $\sum_{i=1}^4 z_i^2=0$. This equation in ${\bf CP}^3$ defines
a copy of ${\bf CP}^1\times {\bf CP}^1$; this is proved
by ``solving'' the equation in terms of $A$'s and $B$'s as
in (\ref{param}). Thus, as claimed, the blowup of
the singularity of ${\bf CP}^2/\Z_2$ gives $\S^2\times \S^2$.

\bigskip\noindent{\it Chiral Operators}
 
Let us now discuss the chiral operators. These operators have
the lowest possible dimension for a given R-charge.
We have assigned the R-charge $1/2$ to each of the $A$'s and
$B$'s. Thus, the lowest possible R-charge for a gauge 
invariant operator is $1$. The corresponding chiral operators, namely
\be 
\label{firstop}
{\rm Tr} A_k B_l \ ,
\ee
have dimension $3/2$ and transform as $(2,2)$ under the 
$SU(2)\times SU(2)$ global symmetry.
In general, we find chiral operators of positive integer R-charge
$n$ and dimension $3n/2$,
\be
\label{genop}
C_L^{ k_1 k_2\ldots k_n} C_R^{ l_1 l_2\ldots l_n}
\Tr A_{k_1} B_{l_1} A_{k_2} B_{l_2} \ldots A_{k_n} B_{l_n}
\ .
\ee
The equations for a critical point of the superpotential 
\be
B_1 A_k B_2 = B_2 A_k B_1\ ,\qquad
A_1 B_l A_2 = A_2 B_l A_1\ ,
\ee
tell us that (modulo descendants)
we can freely permute all $A$'s and all $B$'s in
(\ref{genop}). Thus, $C_L$ and $C_R$ must be completely
symmetric, and
for R-charge $n$ we find chiral operators in the $(n+1,n+1)$
of $SU(2)\times SU(2)$. If we think of ${\rm Tr}\,A_k B_l$ as a vector of
$SO(4)$, which we write $z_i$, 
then these representations are precisely the symmetric 
traceless polynomials in $z_i$ of order $n$. Note that $z_i$ is
the original conifold coordinate, and that we have obtained the
expected wave functions on the conifold. 

The chiral operators we have found are rather analogous to the traceless
symmetric polynomials $\Tr X^{i_1} X^{i_2}\ldots X^{i_n}$ in the
${\cal N}=4$ SYM theory.
Let us recall that, on the supergravity side, such operators were
identified with modes of $h^\alpha_\alpha$ (the trace of the metric
on the compact manifold) and the
four-form gauge potential on ${\bf S}^5$; these
 are described by scalar spherical
harmonics \cite{Kim}. Thus, we expect that
the spectrum of the chiral operators (\ref{genop})
should coincide with the
spectrum of scalar spherical harmonics on $SU(2)\times SU(2)/U(1)$.
Geometrically, this space is a product of two three-spheres
with combined rotations around one of the axes modded out.
Spherical harmonics on the first three-sphere are labeled by
$(J, J_3, I_3)$, where $J_3$ and $I_3$ are two magnetic quantum numbers.
Similarly, the quantum numbers on the second three-sphere are
$(\tilde J, \tilde J_3, \tilde I_3)$. To mod out the $U(1)$ we
impose the constraint 
\be
I_3 + \tilde I_3 = 0\ .
\ee
Then the R-charge is identified with
\be I_3 - \tilde I_3 = 2 I_3
\ ,
\ee
and the remaining quantum numbers $(J, J_3, \tilde J, \tilde J_3)$
label the $SU(2)\times SU(2)$ representations.
Chiral multiplets are obtained from the smallest representation
(with the smallest quadratic Casimir operator) of given R-charge.
For R-charge $n$, this representation is $(n+1, n+1)$.
This agrees with the spectrum of operators
(\ref{genop}) that we found on the field theory side.

\section{Extensions to M theory}

In the preceding sections we presented a duality between Einstein
space compactifications of Type IIB theory and large $N$
field theories on D3-branes placed at conical singularities.
There are obvious extensions of these results to M-theory.

For instance, the geometry $AdS_7\times X_4$, where $X_4$
is a four-dimensional compact Einstein manifold, is created by
placing a large number of M5-branes at a conical singularity of 
$M_6\times Y_5$ where $Y_5$ is a five-dimensional manifold that is
a cone over $X_4$.

Another class of theories, which has more connections 
with old supergravity literature, concerns compactifications of 
eleven-dimensional supergravity on $AdS_4\times X_7$,
where $X_7$ is a seven-dimensional compact Einstein manifold.
Such backgrounds have been  investigated in some detail (for a classic
review, see \cite{Duff}). In the context of the AdS/CFT correspondence,
each of these backgrounds corresponds to a three-dimensional conformal
field theory. This dual theory may be defined as the infrared
limit of the world volume theory on $N$ coincident M2-branes
placed at a conical singularity of $M_3\times Y_8$.
$Y_8$ is a Ricci-flat manifold, a cone over $X_7$.
Its  metric near the singularity
takes the form
\be
\label{eightmetric}
h_{mn} dx^m dx^n = dr^2 + r^2 g_{ij} dx^i dx^j\ , \qquad (i,j= 1, \ldots, 7)
\ ,
\ee
where $r$ is the radial coordinate which vanishes at the singularity,
and $g_{ij}$ is the metric on $X_7$.
The supersymmetry of the resulting three-dimensional theory is
determined by the holonomy of $Y_8$. 
In fact, manifolds $Y_8$ of
$Spin(7)$, $SU(4)$, $Sp(2)$ or smaller holonomy correspond to 
superconformal theories with ${\cal N}=1,2, 3$ or higher supersymmetry,
respectively. This has been recently pointed out also in \cite{ff}.

The cases most analogous to our D3-brane construction correspond
to three-dimensional ${\cal N}=2$ theories and are found by placing
M2-branes at conical singularities of Calabi-Yau four-folds.
Consider, for instance, the non-compact four-fold defined by
\be
\label{newcone}
\sum_{i=1}^5 z_i^2 = 0
\ .
\ee
The singularity is at $z_i=0$. The set of points at unit distance
from the singularity,
\be
\label{unitd}
\sum_{i=1}^5 |z_i|^2 = 1\ ,
\ee
is a coset $SO(5)/SO(3)$. Indeed, the solutions to (\ref{newcone}),
(\ref{unitd}) are obtained by $SO(5)$ rotations of
\be
z_1 = 1/\sqrt 2\ , \qquad z_2 = i/\sqrt 2\ ,
\qquad z_3 = z_4 = z_5=0\ .
\ee
The subgroup of $SO(5)$ that leaves this solution fixed is $SO(3)$
(acting on $z_3,z_4,z_5$), so the space of solutions is $SO(5)/SO(3)$.

The conifold (\ref{newcone}) is thus a cone over the homogeneous
space $SO(5)/SO(3)$.  This homogeneous space is
the example called $V_{5,2}$ in \cite{castell} and in Table
6 of \cite{Duff}.
The isometries of the conifold are 
$SO(5)\times U(1)$; the $U(1)$ acts by $z_i\to \gamma z_i$, and is a group
of R-symmetries (just as for the example we studied in section 3) because
it acts nontrivially on the canonical line bundle of the conifold.
We hope it will be possible to construct a
three-dimensional field theory corresponding
to M2-branes on (\ref{newcone}).

\section*{Acknowledgements}
We are grateful to J. Figueroa--O'Farrill,
S. Gubser, A. Hanany, N. Seiberg, and
M. Strassler  for discussions.  
The work  of I.R.K.  
was supported in part by US Department of Energy 
grant DE-FG02-91ER40671 and 
by the James S. McDonnell
Foundation Grant No. 91-48.  
The work of E.W. is supported in part by NSF grant PHY-9513835.


\end{document}